\documentclass[
	10pt,
	aps,
	prb,
	showpacs,
	twocolumn,
	amssymb,
	amsmath,
	floatfix,
	a4paper,
	citeautoscript,
	superscriptaddress
]{revtex4-1}

\usepackage[usenames]{color}
\usepackage{graphicx}
\usepackage{hyperref}
\usepackage{bbold}

\newcommand{\Vb}{\ensuremath{V_\mathrm{b}}}
\newcommand{\Vg}{\ensuremath{V_\mathrm{g}}}

\newcommand{\DP}{\ensuremath{\Delta\phi}}
\newcommand{\tr}{\ensuremath{\mathrm{Tr}}}

\newcommand{\GL}{\ensuremath{\Gamma_\mathrm{L}}}
\newcommand{\GR}{\ensuremath{\Gamma_\mathrm{R}}}
\newcommand{\Ga}{\ensuremath{\Gamma_\alpha}}

\newcommand{\Grel}{\ensuremath{\Gamma_\mathrm{rel}}}

\newcommand{\gL}{\ensuremath{\gamma_\mathrm{L}}}
\newcommand{\gR}{\ensuremath{\gamma_\mathrm{R}}}
\newcommand{\ga}{\ensuremath{\gamma_\alpha}}

\newcommand{\wL}{\ensuremath{\omega_\mathrm{L}}}
\newcommand{\wR}{\ensuremath{\omega_\mathrm{R}}}
\newcommand{\wa}{\ensuremath{\omega_\alpha}}

\newcommand{\wmm}{\ensuremath{\omega_{m m'}}}
\newcommand{\Xmm}{\ensuremath{F_{m m'}}}

\newcommand{\HLS}{\ensuremath{\hat{H}_\mathrm{LS}}}
\newcommand{\HS}{\ensuremath{\hat{H}_\mathrm{S}}}
\newcommand{\HB}{\ensuremath{\hat{H}_\mathrm{leads}}}
\newcommand{\HT}{\ensuremath{\hat{H}_\mathrm{tun}}}

\newcommand{\LL}{\ensuremath{\mathcal{L}}}
\newcommand{\JP}{\ensuremath{\mathcal{J}^+}_{\rm R}}
\newcommand{\JM}{\ensuremath{\mathcal{J}^-}_{\rm R}}

\newcommand{\RR}{\ensuremath{\boldsymbol{\mathcal{R}}_\mathrm{R}}}
\newcommand{\Ra}{\ensuremath{\boldsymbol{\mathcal{R}}_\alpha}}
\newcommand{\mat}[1]{\ensuremath{\boldsymbol{#1}}}
\newcommand{\at}[2][]{#1|_{#2}}
\newcommand{\Tr}[2]{\tr_{#1}\{#2\}}

\begin{document}

\title{
Interference and shot noise in a degenerate Anderson-Holstein model
}

\author{Andrea Donarini}
\affiliation{Institute for Theoretical Physics, University of Regensburg, 93040 Regensburg, Germany}

\author{Michael Niklas}
\affiliation{Institute for Theoretical Physics, University of Regensburg, 93040 Regensburg, Germany}

\author{Milena Grifoni}
\altaffiliation{E-mail: \href{mailto:milena.grifoni@ur.de}{milena.grifoni@ur.de}}
\affiliation{Institute for Theoretical Physics, University of Regensburg, 93040 Regensburg, Germany}

\date{\today}


\begin{abstract}%
We study the transport properties of an Anderson-Holstein model with orbital degeneracies and a tunneling phase that allows for the formation of dark states. The resulting destructive interference yields a characteristic pattern of positive and negative differential conductance features with enhanced shot noise, without further asymmetry requirements in the coupling to the leads. The transport characteristics are strongly influenced by the Lamb-shift renormalization of the system Hamiltonian. Thus, the electron-vibron coupling cannot be extracted by a simple fit of the current steps to a Poisson distribution. For strong vibronic relaxation, a simpler effective model with analytical solution allows for a better understanding and moreover demonstrates the robustness of the described effects. 
\end{abstract}
\maketitle

\section{Introduction}

Mechanical degrees of freedom can leave clear signatures in the transport characteristics of a nanojunction, as revealed, for example, by elastic and inelastic electron tunneling spectroscopy of single molecule junctions \citep{Smit2002,Qiu2004,Osorio2007,Czap2019,Thomas2019}. The study of mechanical oscillations in nanojunctions extends, nevertheless, far beyond the spectroscopic level. Vibrations correlate with the electronic structure of the molecule \citep{Rolf2019} and the symmetry of the electronic excitation\citep{Pavlicek2013}, they also reveal coherent electron–nuclear coupling \citep{Repp2010}.
Moreover, the mutual influence between the mechanical and the electronic dynamics can range from being a small perturbation to a large one. In the latter case nonperturbative  effects are visible in the Franck-Condon blockade\cite{Koch2006, Leturcq2009, Yar2011} with associated electronic avalanche\citep{Lau2019}, in the bistability even in non-equilibrium conditions \cite{Wilner2013} suppressed, though, by electronic correlation \cite{Souto2018} , in the regular shuttle dynamics\cite{Gorelik1998,Novotny2003} with virtually vanishing shot noise\citep{Novotny2004}, in run away modes \cite{Lue2010,Erpenbeck2018} which ultimately bring to molecular dissociation.   

The Anderson-Holstein model\citep{Anderson1961,Holstein1959} (AHM) is the minimal model for the description of vibronic effects in an interacting nanojunction. It consists of a {spinful} interacting level coupled to a vibrational mode and to non interacting electrodes. Despite its simplicity, AHM comprises several regimes defined by the mutual relations among its four energy scales: the charging energy $U$, the vibronic excitation energy  $\hbar \omega$, the electron-vibron coupling $\lambda$,  and the level broadening $\hbar \Gamma$ caused by the coupling of the nanojunction to the electrodes. For example, the study of the shot noise and full counting statistics for an AHM\citep{Seoane-Souto2014} has revealed electronic avalanches  dynamics\citep{Koch2005,Koch2006}  and hysteretic or switching behavior \citep{Galperin2005,Ryndyk2008,Pistolesi2008,Donarini2012b} in the Frank-Condon blockade, sub-Poissonian noise due to relaxation \citep{Haupt2006} and absorption side bands in the Coulomb blockade regime \citep{Luffe2008}. 

Suspended carbon nanotubes (CNTs) are ideal systems to study vibronic effects in interacting nanojunctions and have stimulated an intense research activity. Several transport experiments in these systems \citep{Sapmaz2006,Leturcq2009,Huettel2009} indeed show  clear indications for vibrational excitations, with characteristic Poisson distributed steps, as predicted for an AHM with thermally equilibrated vibrons \citep{Braig2003}. Suspended CNTs are also excellent electromechanical resonators \citep{Babic2003, Sazonova2004, Lefevre2005, Steele2009, Huttel2009,Eichler2011} with quality factors up to $10^6$\citep{Island2012,Moser2014}. Even tailoring of the electron-vibron coupling with gate voltages \cite{Benyamini2014,Weber2015}, electronic injection position\citep{Cavaliere2010,Cavaliere2010b,Ziani2011} or  magnetic fields\citep{Stiller2018} has been demonstrated. 

The transport characteristics of CNT quantum dots still present, though, puzzling additional features, not captured by the {simple} AHM, as the alternating positive and negative differential conductance (PDC/NDC) at the vibrational side bands \citep{Sapmaz2006,Leturcq2009}. The alternance of PDC/NDC at bosonic side bands has so far been attributed to large asymmetries in the tunneling coupling of the CNT levels to the leads \citep{Cavaliere2004,Zazunov2006,Shen2007}, combined with level-splittings of the order of half of the vibrational energy \citep{Yar2011, Leturcq2009, Cavaliere2010, Cavaliere2010b} or to image-charge effects \citep{Perfetto2013}. An extension of the Anderson Holstein model  including the electron-vibron coupling also in the exchange parameter has been proposed to explain the switching in the coupling to phonons in a suspended CNT quantum dot \cite{Weber2015}. 
All these models however fail to explain the strong drop in current at large bias voltages, as they instead predict a current saturation. Additionally, in the theoretical {prediction} the NDC has  the form of clearly defined steps \citep{Cavaliere2010,Yar2011}, contrasting with the much smoother behavior exhibited by the measurements \citep{Sapmaz2006,Leturcq2009,Stiller2018}.

Recent experiments on CNTs have demonstrated quantum interference as a further source of NDC \citep{Donarini2019}, which yields to coherent population trapping and dark state formation. Similar effects have been reported for triple dots\citep{Michaelis2006, Donarini2010, Niklas2017}, semiconductor nanowires quantum dots\citep{Nilsson2010,Karlstroem2011} and single molecule junctions\citep{Donarini2009,Donarini2012}. Necessary condition for the occurrence of such phenomenon is the presence of two orbitally degenerate states, supported in the CNT by the valley degree of freedom. The control of such a degree of freedom goes under the name of valleytronics. This concept has been recently extended to the one of flavortronics\cite{Maurer2020} for systems with higher system degeneracies.   

Motivated by these results and by the puzzling anomalies in the vibrational fingerprints of the suspended CNTs, we study in this paper a degenerate AHM, which combines interference blocking and vibrational excitations. The obtained current voltage characteristics qualitatively reproduce the ones of the experiments\citep{Sapmaz2006,Leturcq2009} and give for them an alternative explanation which is based on electrical dark states and it does not require orbital asymmetries nor carefully tuned level splittings. A crucial role is played instead by virtual electronic fluctuations which produce a Lamb shift-like renormalization of the system Hamiltonian. As such, the Bloch vector associated to the orbitally degenerate states precesses around an exchange field \citep{Braun2004,Donarini2009,Schultz2010,Hell2015}, modulating in this way the degree of interference blocking and thus the current and the shot noise of the nanojunction. The interplay of such pseudospin precession with the vibrational degree of freedom and the  fingerprints left in the current and noise of a degenerate AHM represent the main topic of this work. The theory developed here is not restricted to CNTs. It {rather} naturally applies to a whole class of single molecule junctions characterized by symmetry protected orbital degeneracies \citep{Park2000,Donarini2009,Donarini2012,Lau2019}. 
 
The paper is structured as follows. In Sec. II we present the degenerate Anderson-Holstein model and the relevant dynnamical equations for the reduced density operator. In Sec. III we present numerical results for the current and the Fano factors, where interference effects are clearly visible in their associated stability diagrams. We focus in Sec. IV on the regime of strong relaxation, which allows us an analytical treatment of the dynamics and a clear physical interpretation of the results of the Sec. III. Finally, the robustness of the interference features against various perturbation is discussed in Sec. V and we draw our conclusions in Sec. VI.

\section{Degenerate Anderson-Holstein model}
\begin{figure}[t!]
	\centering
	\includegraphics{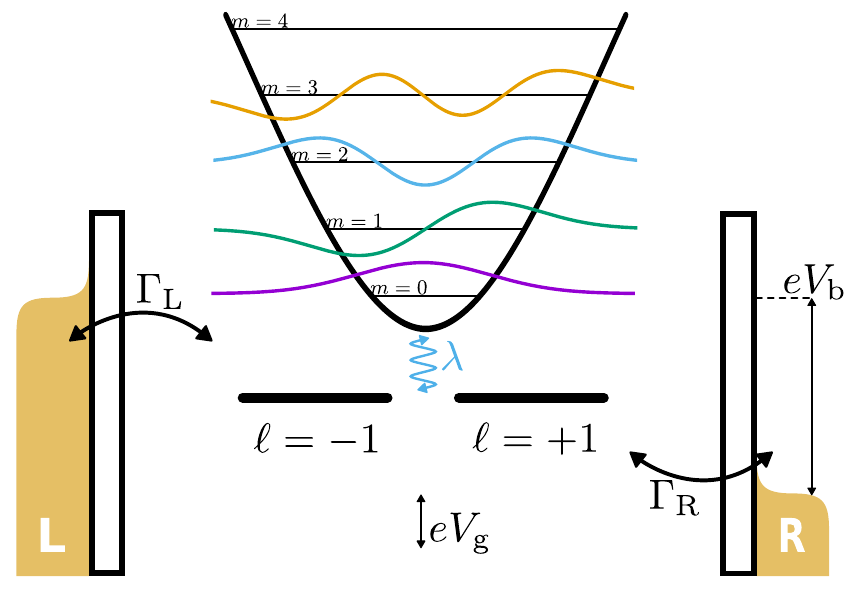}
	\caption{Anderson-Holstein meodel with orbital degeneracies coupled to two leads. The electron-vibron coupling, {characterized by the strength} $\lambda$, allows for vibronic excitations induced by the tunneling of electrons between the leads and the degenerate levels. The current through the system is controlled by sweeping the bias and the gate voltages, respectively $V_{\rm b}$ and $V_{\rm g}$.}
	\label{fig:model}
\end{figure}

The Anderson-Holstein model (AHM) is a standard choice for the simultaneous description of electron-electron and electron-vibron interaction in nanodevices \cite{Boese2001,Hettler2002,Braig2003,Novotny2003,Cizek2004,Galperin2005,Kaat2005,Yar2011}  and several techniques have been applied to study it, ranging from a master equation approach \cite{Koch2006} to the more sofisticated multilayer multiconfiguration time-dependent Hartree \cite{Wilner2013}, the self-consistent diagrammatic approximation \cite{Souto2018} or the real-time path-integral Monte Carlo approach \cite{Muehlbacher2008} . Here we consider the AHM to a quantum dot with orbital degeneracy, such to incorporate also quantum interference. We thus consider the Hamiltonian $\hat{H}=\hat{H}_\mathrm{S} + \hat{H}_\mathrm{leads} + \hat{H}_\mathrm{tun}$ for a quantum dot (QD) with spin $\sigma \in \{\uparrow,\downarrow\}$ and orbital $\ell \in \{+,-\}$ degrees of freedom coupled to two leads, see Fig.~\ref{fig:model}. 

All the four single particle states in the dot are degenerate with the onsite energy $\varepsilon = e\Vg$.  Coupling to a phononic mode results in the system Hamiltonian
\begin{align}
\label{eq:H_AndersonHolstein}
	\HS
	&=
	\varepsilon \hat{N}
	+ \frac{U}{2}\hat{N}\left(\hat{N}-1\right)
	\nonumber \\
	&\quad + \hbar\omega\left(a^\dagger a+\frac{1}{2}\right)
	+ \lambda \hat{N} \left(a^\dagger + a\right)
	,
\end{align}
where $\hat{N} = \sum_{\ell\sigma} d^\dagger_{\ell\sigma} d_{\ell\sigma}$. Here, $d^{(\dagger)}_{\ell\sigma}$ destroys (creates) an electron in the QD.
The leads are described as free Fermi gases with $\HB = \sum_{\alpha k \sigma} \varepsilon_{\alpha k} c^\dagger_{\alpha k \sigma} c_{\alpha k \sigma}$, where $c^{(\dagger)}_{\alpha k \sigma}$ destroys (creates) an electron in lead $\alpha$ with momentum $k$ and spin $\sigma$. The nanojunction is brought out of equilibrium by a voltage drop $\Vb$ which enters the chemical potentials $\mu_\mathrm{L} = \eta e\Vb$ and $\mu_\mathrm{R} = (\eta-1)e\Vb$ of the electrodes. Additionally, the system is kept at the temperature $T$ and the tunneling rates are proportional to the Fermi functions $f^+_\alpha(x) = 1/\left(1 + \exp[(x-\mu_\alpha)/(k_BT)]\right)$ and $f^-_\alpha(x) = 1-f^+_\alpha(x)$.

The diagonalization of $\HS$ is commonly obtained via the polaron transformation $e^{\hat{P}} \HS e^{-\hat{P}}$ with $\hat{P} = \frac{\lambda}{\hbar\omega}\hat{N}(a^\dagger-a)$\citep{Mahan2000} which effectively decouples the electronic and the vibronic degrees of freedom. The resulting eigenvectors and eigenvalues read
\begin{align}
\label{eq:eigenvectors}
	\vert NmSS_zL_z\rangle
	&=
	e^{-\hat{P}}\vert NSS_zL_zm\rangle_0,
	\\
	E_{Nm}
	&=
	\tilde{\varepsilon}N + \frac{\tilde{U}}{2}N(N-1) + \hbar\omega\left(m+\frac{1}{2}\right),
\end{align}
where the eigenvectors Eq.~\eqref{eq:eigenvectors} are obtained by polaronic transformation of the factorized states $\vert NSS_zL_zm\rangle_0 = \vert NSS_zL_z\rangle_0 \otimes \vert m\rangle$. The quantum numbers which fully characterize the electronic many-body states of the system are the total occupation $N$, the total spin $S, S_z$, the imbalance between orbital occupations $L_z = \sum_{\sigma} (n_{+,\sigma} - n_{-,\sigma})$, and finally the number of vibronic excitations $m$. The eigenenergies depend on the total particle number $N$, the normalized energies $\tilde{\varepsilon} = \varepsilon-\frac{\lambda^2}{\hbar\omega}$, $\tilde{U} = U-\frac{2\lambda^2}{\hbar\omega}$ and on the vibronic energy $\hbar\omega$ with its excitation level $m$. 
Hence, the spectrum consists of equidistant (bosonic) replicas of the electronic energy ground state. No signature of the bare single particle energy and charging energy can be observed in the tunneling spectroscopy. We therefore will drop in the following the tildes and return to $U$ and $\varepsilon$ for the measurable charging and single particle energies.

The tunneling Hamiltonian $\HT = \sum_{\alpha k \ell \sigma} t_\alpha e^{i \ell \phi_\alpha} c^\dagger_{\alpha k \sigma} d_{\ell\sigma} + \mathrm{h.c.}$ incorporates an orbital and lead dependent phase $\ell\phi_\alpha$ which ultimately enters in the tunneling rate matrix for the lead $\alpha$
\begin{equation}
\label{eq:Gmat}
	\boldsymbol{\Gamma}_\alpha = \Ga\Ra,
\end{equation}
with the bare tunneling rate $\Ga = 2\pi \vert t_\alpha \vert^2 D_\alpha/\hbar$ proportional to the density of states at the Fermi level $D_\alpha$, and the coherence matrix in the orbital basis
\begin{equation}
\label{eq:Rmat}
	\boldsymbol{\mathcal{R}}_\alpha
	=
	\begin{pmatrix}
		1 & a_\alpha e^{2i\phi_\alpha} \\
		a_\alpha e^{-2i\phi_\alpha} & b_\alpha^2
	\end{pmatrix}
	,
\end{equation}
with $|a_\alpha| < |b_\alpha|$, $b_\alpha \in \mathbb{R}$ ensuring the positivity of the rate matrix $\boldsymbol{\Ga}$.

The coherence matrix is the central object of our model as it accounts for the possibility of orbital interference \citep{Karlstroem2011,Niklas2017,Donarini2019}. In the following we assume equal coupling of the orbitals to the lead $\alpha$, such that $b_\alpha = 1$. Further, in the spirit of the surface $\Gamma$-point approximation \citep{Donarini2019}  we set $a_\alpha = 1$.

We concentrate on the weak tunneling limit $\hbar\Ga \ll k_B T,U$ and describe the dynamics of the system using a master equation for the reduced density matrix $\hat{\varrho}  = {\rm Tr}_{\rm leads}\{\hat{\varrho} _{\rm tot}\}.$ 
Such a matrix is block-diagonal in particle number and spin, due to conservation of these quantities in the total Hamiltonian. Moreover, in the limit of fast vibrations, $\omega \gg \Ga$, the secular approximation \citep{Blum2012} also ensures the vanishing of coherences among states with different vibronic excitation. Finally, due to the spin isotropy of the leads, we use the Wigner-Eckart theorem and integrate out $S_z$. 
We thus define, conveniently,  the matrix element in the ${NmS}$ block as $\varrho^{NmS}_{L_z L_z'}=\sum_{S_z}{\rm Tr}\{\hat{\varrho}\vert N m S S_z L_z \rangle\langle N m S S_z L_z' \vert\}$. To avoid unnecessary complications, we restrict our analysis to the transitions between states with zero and one electron on the dot. This regime can be obtained, in the limit of large charging energy $U \gg \hbar\Gamma, \hbar\omega$, by tuning the gate and bias voltage in the vicinity of the zero to one particle resonance. To leading order in the tunneling coupling, we obtain the following generalized master equation (GME):
\begin{align}
\label{eq:master_equation}
	\dot{\varrho}^{0m0}
	&=
	-4\sum\limits_{\alpha m'}
	\Ga
	\Xmm
	f^+_\alpha\Bigl(\varepsilon-\wmm\Bigr)
	\varrho^{0m0}
	\nonumber \\
	&\quad+
	\sum\limits_{\alpha m'}
	\Ga
	\Xmm
	f^-_\alpha\Bigl(\varepsilon-\wmm\Bigr)
	\tr \left\{\Ra \mat{\varrho}^{1m'\frac{1}{2}} \right\}
	, \nonumber \\	
	\dot{\mat{\varrho}}^{1m\frac{1}{2}}
	&=
	-\frac{i}{\hbar}
	\left[\mat{H}_{\rm LS}^{1m\frac{1}{2}},\mat{\varrho}^{1m\frac{1}{2}}\right]
	\nonumber \\
	&\quad
	-\frac{1}{2}
	\sum\limits_{\alpha m'}
	\Ga
	\Xmm
	f^-_\alpha\Bigl(\varepsilon+\wmm\Bigr)
	\left\{\Ra, \mat{\varrho}^{1m\frac{1}{2}}\right\}
	\nonumber \\
	&\quad+
	2\sum\limits_{\alpha m'}
	\Ga
	\Xmm
	f^+_\alpha\Bigl(\varepsilon+\wmm\Bigr)
	\Ra \varrho^{0m'0}
	,
\end{align}
where $[\mat{A},\mat{B}]$ and $\{\mat{A},\mat{B}\}$ indicates the commutator ($\mat{A}\mat{B}-\mat{B}\mat{A}$) and the anticommutator ($\mat{A}\mat{B}+\mat{B}\mat{A}$) relations, respectively. Eq.\eqref{eq:master_equation} describes in and out tunneling processes between the quantum dot and the leads, as well as the pseudospin precession due to virtual charge fluctuations involving degenerate orbitals. This {latter} aspect of the system dynamics is captured by the Lamb-shift Hamiltonian, which explicitly reads:
\begin{align}
\label{eq:H_LS}
	\mat{H}_{\rm LS}^{1m\frac{1}{2}}
	&=
	\frac{\hbar}{2\pi}
	\sum\limits_{\alpha m'}
	\Ga
	\Ra
	\Xmm
	\nonumber \\
	&\quad
	\times
	\left[
		p_\alpha\Bigl(\varepsilon+\wmm\Bigr)
		-
		p_\alpha\Bigl(\varepsilon+U-\wmm\Bigr)
	\right]
	,
\end{align}
with $\wmm = \hbar\omega(m-m')$, and the function $p_\alpha(x) = -\operatorname{Re}\Psi \left[ 1/2 + i(x-\mu_\alpha)/(2\pi k_BT) \right]$ defined via the digamma function $\Psi$. 
Here, $\hat{\varrho}^{0m0}$ is the probability of finding the quantum dot at the same time empty and oscillating with $m$ vibronic excitations. The one particle component $\mat{\varrho}^{1m\frac{1}{2}}$ is instead a $2\times2$ matrix due to the orbital degree of freedom and, beyond the occupation probability, it carries information on the pseudospin, whose components are calculated as

\begin{equation}
    T_i = \frac{\sum_{m} \Tr{}{\mat{\rho}^{1m\frac{1}{2}}\mat{\sigma}^i}}
    {\sum_{m} \Tr{}{ \mat{\rho}^{1m\frac{1}{2}}}},
\end{equation}
with $\mat{\sigma}^i$ the $i$th Pauli matrix. The Franck-Condon factors $\Xmm$ account for the wave function overlap between the initial and final vibronic state. They are defined as \citep{Yar2011}
\begin{align}
\label{eq:FC_factors}
	\Xmm
	&=
	\vert \langle m \vert e^{-\tfrac{\lambda}{\hbar \omega}(a^\dagger - a)} \vert m' \rangle \vert^2
	\nonumber \\
	&=
	e^{-g}
	g^{\vert m'-m \vert}
	\left(\frac{m!}{m'!}\right)^{\mathrm{sgn}(m'-m)}
	\nonumber \\
	&\qquad
	\times
	\left[L^{\vert m'-m \vert}_{\min(m,m')}(g)\right]^2
	,
\end{align}
with the associated Laguerre polynomials $L^{k}_{n}(x)$ and $g = (\lambda/\hbar\omega)^2$ the dimensionless coupling constant. 

The model Hamiltonian Eq.~\eqref{eq:H_AndersonHolstein} and the equation of motion Eq.~\eqref{eq:master_equation} capture the interplay of two effects: vibron assisted tunneling and interference blocking. On one hand, the sequential tunneling ensured by the small tunneling coupling is modulated by the Franck-Condon factors $\Xmm$. The dimensionless electron vibron coupling $g = (\lambda/\hbar\omega)^2$ is a measure of the relevance of the vibron assisted tunneling phenomena. The coupling $g \approx 3$ considered here gives already well defined Stokes side peaks in the tunneling spectroscopy. Its moderate value, though, keeps the dynamics far from the bistable Franck-Condon blockade regime with its avalanche dynamics\citep{Koch2005,Koch2006} characterized by giant noise. On the other hand, the phase difference $\phi_L \neq \phi_R$ in the electronic rate matrices [see Eq.~\eqref{eq:Gmat}] connecting degenerate states supports the formation of dark states and the pseudospin precession. 

The interplay of vibron assisted tunneling and interference blocking is at the origin of the transport phenomena described in the next section.     

\section{Current and noise}
The calculation of the transport properties moves from the Markovian limit of full counting statistics for electron transport \citep{Flindt2005,Marcos2010,Emary2012}. A central role is played in this formalism by the generalized reduced density operator $\hat{R}_\alpha(\chi) = \tr_{\rm leads}\{e^{i\chi \hat N_\alpha}\hat{\varrho}_{\rm tot}\}$ which contains the counting field $\chi$. In {terms} of $\hat{R}_\alpha$ we define the current cumulants $c_{\alpha,n}$:

\begin{equation}
 \label{eq:Curr_cumulant}
c_{\alpha,n} = \frac{d}{dt}\left(\frac{\partial}{\partial i\chi}\right)^n \ln \tr\{\hat{R}_\alpha(\chi)\}\at[\Big]{\chi = 0}
\end{equation}

The first two current cumulants are, by rescaling with the electronic charge $e$, the current and current noise, 

\begin{equation}
\label{eq:Current_Noise}
\begin{split}
    c_{\alpha,1} &= \frac{d}{dt}\langle \hat{N}_\alpha \rangle = \frac{I_\alpha}{e},\\
    c_{\alpha,2} &= \frac{d}{dt} \left(\langle\hat{N}^2_\alpha \rangle - \langle\hat{N}_\alpha \rangle^2\right) = \frac{S_\alpha}{e^2}, 
\end{split}    
\end{equation}
where $\langle \bullet \rangle = \tr_{\rm tot}\{\bullet\hat{\varrho}_{\rm tot}\}$ indicates the quantum mechanical average. Both $I_\alpha$ and $S_\alpha$ depend on time and are associated to a specific lead. In the stationary limit, though, $t \to \infty$, $I_{\rm L}$ = -$I_{\rm R}$ and $S_{\rm L} = S_{\rm R}$. Clearly, the calculation of the current cumulants requires to know the dynamics of $\hat{R}_\alpha(\chi)$. If the equations of motion Eq.~\eqref{eq:master_equation} define a Liouville superoperator acting on the reduced density matrix $\mathcal{L}\hat{\varrho} = \dot{\hat{\varrho}}$,  the equation of motion for the operator $\hat{R}_\alpha(\chi)$ reads:

\begin{equation}
\label{eq:GME_FCS}
    \dot{\hat{R}}_\alpha(\chi)= [\mathcal{L} + \mathcal{J}_\alpha(\chi)]\hat{R}_\alpha(\chi)
\end{equation} 
where we have introduced the superoperator
\begin{equation}
    \mathcal{J}_{\alpha}(\chi) = (e^{i\chi}-1)\mathcal{J}^+_\alpha + (e^{-i\chi}-1)\mathcal{J}^-_\alpha
\end{equation}
expressed in terms of the counting field $\chi$ and the forward and backward jump superoperators $\mathcal{J}^\pm_\alpha$.
The latter are:

\begin{equation}
    \begin{split}
        \mathcal{J}^+_\alpha &= \sum_{\ell \ell'\sigma}(\boldsymbol{\Gamma}_\alpha)_{\ell\ell'}
        \mathcal{D}_{\ell\sigma,+} f^-_\alpha(i\hbar \mathcal{L}_{\rm S})
        \mathcal{D}^\dagger_{\ell'\sigma,-}\,,\\
        \mathcal{J}^-_\alpha &= \sum_{\ell \ell'\sigma}(\boldsymbol{\Gamma}_\alpha)_{\ell\ell'}
        \mathcal{D}_{\ell\sigma,-} f^+_\alpha(i\hbar \mathcal{L}_{\rm S})
        \mathcal{D}^\dagger_{\ell'\sigma,+}.\\
    \end{split}
\end{equation}
The last two equations contain the system Liouvillean $\mathcal{L}_{\rm S} = -i/\hbar[\hat{H}_{\rm S},\bullet]$, together with the superoperators $\mathcal{D}_{\ell\sigma,\pm}$ and $\mathcal{D}^\dagger_{\ell\sigma,\pm}$, which act on a generic operator $\hat{O}$ as:
\begin{equation}
    \begin{split}
       \mathcal{D}_{\ell\sigma,+}\hat{O} &= d_{\ell \sigma}\hat{O}\,,\quad 
       \mathcal{D}_{\ell\sigma,-}\hat{O} = \hat{O}d_{\ell \sigma}\,;\\
       \mathcal{D}_{\ell\sigma,+}^\dagger\hat{O} &= d_{\ell \sigma}^\dagger\hat{O}\,,\quad 
       \mathcal{D}_{\ell\sigma,-}^\dagger\hat{O} = \hat{O}d_{\ell \sigma}^\dagger\,.\\
    \end{split}
\end{equation}
When projecting on the eigenbasis of the system consistently with approximations introduced in Eq.~\eqref{eq:master_equation}, we obtain, for the (right lead) jump operators: 

\begin{align}
\label{eq:current_operators}
	\left[\JP \hat{\varrho}\right]^{0m0}
	&=
	\sum\limits_{m'}
	\GR
	\Xmm
	f^-_\mathrm{R}\Bigl(\varepsilon-\wmm\Bigr)
	\tr \left\{\RR \mat{\varrho}^{1m'\frac{1}{2}} \right\}
	, \nonumber \\
	\left[\JP \hat{\varrho}\right]^{1m\frac{1}{2}}
	&=
	0
	, \nonumber \\
	\left[\JM \hat{\varrho}\right]^{0m0}
	&=
	0
	, \nonumber \\
	\left[\JM \hat{\varrho}\right]^{1m\frac{1}{2}}
	&=
	2\sum\limits_{m'}
	\GR
	\Xmm
	f^+_\mathrm{R}\Bigl(\varepsilon+\wmm\Bigr)
	\RR
	\hat{\varrho}^{0m'0}.
\end{align}
Being proportional to the tunneling rates, also the jump operators are dressed with the Franck-Condon factors $\Xmm$, due to the vibronic degree of freedom. 
Current $I$ and shot noise $S$ are calculated by iteratively evaluating the following hierarchy of equations for the stationary generalized reduced density operator $\hat{R}_\infty = \hat{R}(t\to\infty)$ \citep{Benito2016},
\begin{equation}
\label{eq:Hierarchy}
\begin{split}
	\LL \hat{\varrho}_\infty 
	&=
	0
	, \\	
	I
	&=
	e\tr\left\{(\JP-\JM)\hat{\varrho}_\infty\right\}
	, \\
	\LL \hat{X}_{1\infty}
	&=
	-\left( \JP - \JM - I/e \right)\hat{\varrho}_\infty
	, \\
	S
	&=
	e^2
	\tr \left\{ 2(\JP-\JM)\hat{X}_{1\infty} + (\JP+\JM)\hat{\varrho}_\infty \right\}
	,
	\end{split}
\end{equation}
with the negative electrical charge $e$. In Eqs.~\eqref{eq:Hierarchy} we have introduced the operator $\hat{X}_{1\infty}$, defined as 
the case $n = 1$ of the more general Taylor coefficient
\begin{equation}
    \hat{X}_{n\infty} = \lim_{t \to \infty} \left(\frac{\partial}{\partial i\chi}\right)^n\frac{\hat{R}}{\tr\{\hat{R}\}}\at[\Big]{\chi = 0},
\end{equation}
We further notice that $\hat{X}_{0\infty} = \hat{\varrho}_\infty$. In general, the full counting statistics is obtained by the recursive solution of the set of coupled equations \cite{Benito2016}:

\begin{equation}
    \begin{split}
        c_{\alpha,k} &= \sum_{k' = 0}^{k-1}\binom{k}{k'}\tr\{(\mathcal{J}^+_\alpha + (-1)^{k-k'}\mathcal{J}^-_\alpha) \hat{X}_k'\}\\
        \dot{\hat{X}}_k &= \mathcal{L}\hat{X}_k\! +\! \sum_{k'= 0}^{k-1}\binom{k}{k'}(\mathcal{J}^+_\alpha + (-1)^{k-k'}\mathcal{J}^-_\alpha - c_{k-k'})\hat{X}_{k'}.
    \end{split}
\end{equation} 

If we consider the full time dependence of the current cumulants, the finite frequency noise can be calculated \cite{Flindt2008}. An example of such calculation for  the AHM is given in \cite{Stadler2018}.

\begin{figure}
	\centering
	\includegraphics{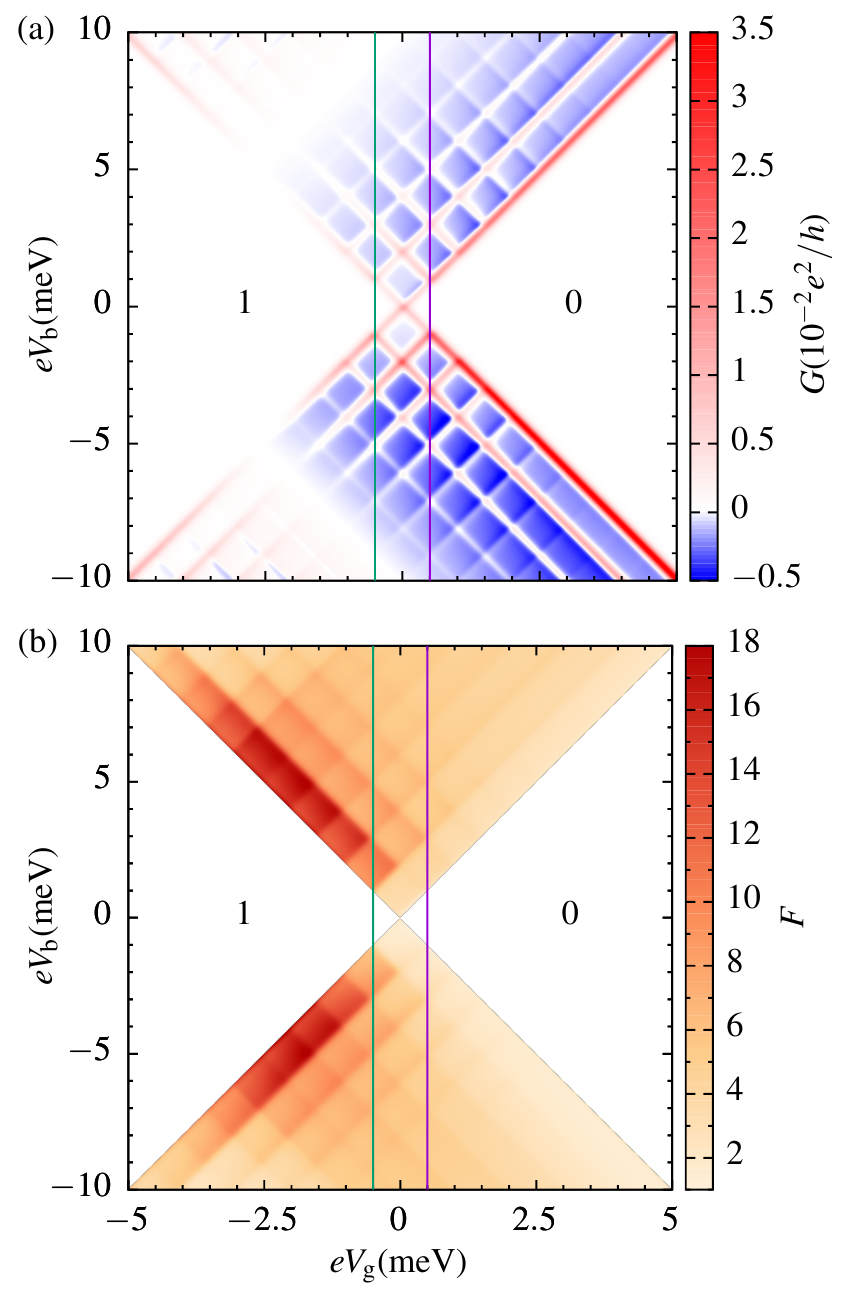}
	\caption{Stability diagram of (a) differential conductance and (b) Fano factor. The electron numbers on the QD are indicated in the Coulomb blockade regions, where the Fano factor is not displayed.
	Parameters are $U=10$\,meV, $\hbar \GL=10\,\mu$eV, $\hbar\GR=15\,\mu$eV, $e\Vg=1$\,meV, $\hbar\omega=1$\,meV, $\lambda=1.8$\,meV, $\eta=0.5$, $k_{\rm B}T=50\,\mu$eV, $\phi_{\rm R} - \phi_{\rm L} = 0.25\pi$, $\hbar\Grel = 0$.
	}
	\label{fig:stab_diag}
\end{figure}

\begin{figure}
	\centering
	\includegraphics{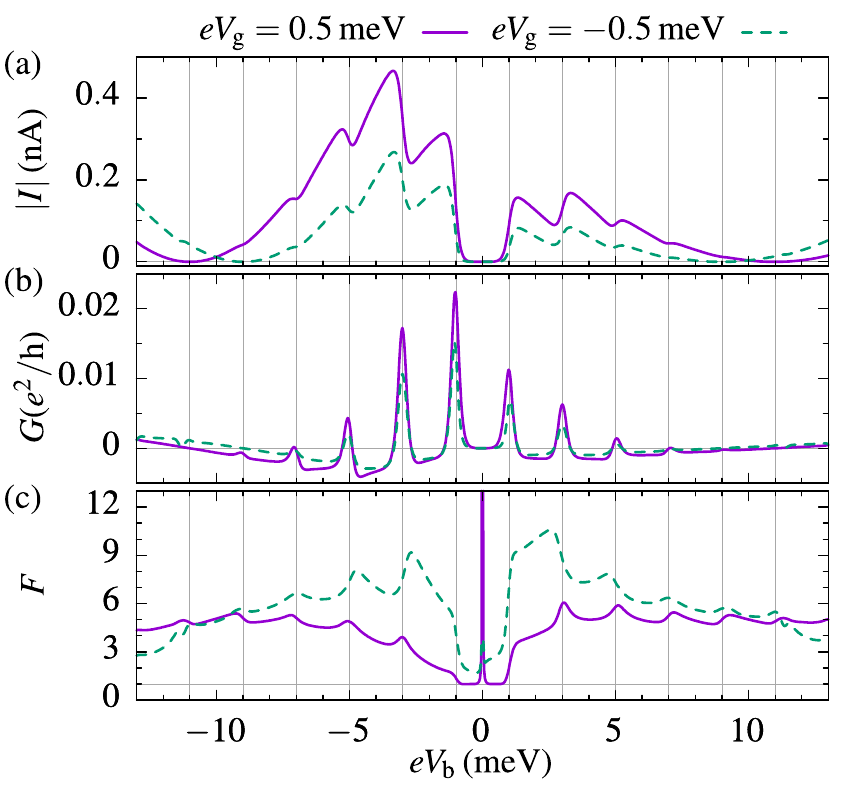}
	\caption{Bias traces at $eVg=\pm0.5$\,meV, as indicated by the vertical lines in Fig.~\ref{fig:stab_diag}.
		(a) Current, (b) differential conductance and (c) Fano factor.}
	\label{fig:bias_trace}
\end{figure}

We concentrate here on the zero frequency components. We employ the Fano factor $F = \vert S/eI \vert$ as dimensionless measure for the noise with respect to its Poissonian value $F = 1$.
In Fig.~\ref{fig:stab_diag} we show the stability diagram of the differential conductance, $dI/dV_{\rm b}$ and the Fano factor. 
In the Coulomb blockade regions (indicated in white) the particle number is fixed and no current flows through the nanojunction. The Fano factor is not calculated in this parameter regime in which the cotunneling contribution, discarded in this work, is expected to dominate the transport characteristics. Current and differential conductance also vanish at the antiresonance conditions $e\Vg + \eta e\Vb = -U/2$ and $e\Vg + (\eta-1) e\Vb = -U/2$, due to interference blocking. Such a condition is obtained when the Lamb shift correction only contains the contribution of the drain. The pseudospin precession dynamics is hindered and the blocking is perfect\cite{Donarini2009}.  Interestingly, this antiresonance only results in a moderate super Poissonian shot noise ($F\approx6$)  as compared to the large values of the Fano factor ($F\approx20$) which can be observed at the edge of the one electron Coulomb diamond. 
For a better quantitative analysis we show cuts through the stability diagram at $e\Vg = \pm 0.5$\,meV in Fig.~\ref{fig:bias_trace}. Here, for completeness, we kept the values in the Coulomb blockade regions, not displayed in the stability diagram of  Fig.~\ref{fig:stab_diag}(b), and one at least appreciates the divergent Fano factor at small biases $\Vb \to 0$ associated to the thermal Johnson-Nyquist noise. 

 The interplay between the electronic interference and the vibron assisted tunneling is clearly visible in the I-V characteristics of Fig.~\ref{fig:bias_trace}(a). 
 
Interference blocking associated to the valley degree of freedom sets the smooth behavior of the current, which, for positive bias, raises at the resonance condition $e\Vg + \eta e\Vb = 0$, and decreases then until complete interference blocking at the antiresonance condition $e\Vg + \eta e\Vb = -U/2$. The smooth transition to the complete blocking is set by the (large) energy scale $U$ and by the shape of the digamma function. The physics of the junction is governed here by the precession between the blocking and the conducting states induced by virtual charge fluctuations\cite{Donarini2010}. 

The smaller scale $\hbar \omega$ which charaterizes the I-V characteristics is given, instead, by the electron-vibron coupling. Vertical gray lines highlight the bias at which new vibronic excitations enter the transport window. As soon as the $m$th vibronic excitation enters the bias window, the current raises with a step, which is related to the sum of Franck-Condon factors $\sum_n F_{n,n+m}$. Due to the moderate value of the dimensionless electron-vibron coupling $g = 3.24$, the steps are smaller at higher biases, since, independent of the initial state $n$, $\lim_{m\to \infty } F_{n,n+m} = 0$.   

If the vibrons quickly equilibrate between two consecutive tunneling events at a temperature $k_{\rm B} T \ll \hbar \omega$, the step height in the I-V curve follows a Poissonian distribution. Indeed, the Franck-Condon factors for transitions involving the ground state  are $F_{0,m} \propto e^{-g}g^m/m!$ \cite{Braig2003}. More concretely, the fast relaxation excludes a feedback of the current on the vibronic distribution. Thus, each additional vibron contributes independently to the current and, if energetically available, is excited with probability $g$ (independent boson model). The analysis of the step height distribution allows, in this approximation, to extract from the I-V characteristics the value of the dimensionless electron-vibron coupling $g = (\lambda/\hbar\omega)^2$.  Such a procedure, though, would even qualitatively fail for the degenerate Anderson-Holstein model. As seen in Fig. \ref{fig:bias_trace}, for example, the third current step is lower than the second one in clear contradiction to the result $g/2 = 1.62$ for their ratio assuming a Poisson distribution with mean value $g = 3.24$. 

In Fig.~\ref{fig:bias_trace}(b) we show the differential conductance through the system. The presence of degenerate interfering states explains the appearance, for every vibration assisted tunneling peak, of a corresponding negative differential conductance (NDC) valley. Such current peaks with an alternating differential conductance have been observed in the current voltage characteristics of suspended carbon nanotubes \cite{Sapmaz2003,Sapmaz2006,Leturcq2009} and have been attributed to strongly asymmetric coupling of quasi degenerate  electronic states \cite{Cavaliere2010,Cavaliere2010b,Yar2011} or, in scanning tunneling microscopy configurations \cite{LeRoy2004}, to a half shuttle dynamics \cite{Zazunov2006}. 

A different tunneling coupling of the nanotube to the two leads is not surprising and can be easily and independently  verified (e.g. by analyzing the strength or the slope of the corresponding resonant lines in the stability diagram). Instead, an asymmetry between the valley degrees of freedom is more difficult to imagine. Both wave functions are, in fact, essentially uniform along the nanotube waist. Moreover, an {\it ad hoc} matching of the energy scales of the vibronic excitations with the one of the electronic level splitting within each shell must be assumed, in order to reproduce the regular alternance of positive and negative differential conductance of the experimental results \cite{Sapmaz2006,Leturcq2009}. 
Within the interference blocking picture, instead, neither an energy scale matching nor tunneling coupling asymmetry are required. The quasidegeneracy of the electronic states and the local tunneling necessary for dark state formation \cite{Donarini2019} are enough, independent of the vibronic energy. 

The interplay between orbital degeneracy, interference and vibronic excitation has already been investigated \cite{Schultz2010, Salhani2017}, although without a systematic study of the current noise. The latter, as shown in Fig.~\ref{fig:bias_trace}(c) shows maxima in correspondence of the minima of the current, i.e. every time a new vibronic copy of the interference blocking state hinders the current, thus provoking further bunching. In this spirit one also understands why the maximum of the Fano factor is reached in correspondence to the first NDC and not at the complete current blocking (antiresonance condition). With the charging energy taken much larger that the vibronic one, the complete interference blocking can only be reached when several vibronic channels have already entered the transport window. These multiply the possible transport paths of the electron through the junction, thus hindering the bunching dynamics. 
 
A bias larger than $\hbar\omega$ allows {one} to indefinitely climb up the regular ladder of the vibronic spectrum via a sequence of energetically allowed transition: for example with every tunneling event depositing a quantum of energy in the mechanical degree of freedom. In the model presented so far, we allowed the vibron to relax uniquely via its coupling to the dissipative electronic dynamics. Under these conditions the results must be carefully checked upon truncation of the number of excitations. We retain up to $30$ excitations, with a cut-off threshold of $10^{-10}$ for the population of the most excited state.

\section{Strong vibrational relaxation}

The mechanical oscillations in a nanojunction are not damped only by the electronic reservoirs. Both the intrinsic anharmonicities of the internal vibrational modes, as well as the direct coupling to the mechanical degrees of freedom of the environment represent further sources of dissipation.  In the absence of external driving, we thus expect, eventually, thermal  equilibration. For simplicity, we refrain from formulating a microscopic model for the mechanical dissipation. We introduce instead in Eqs.~\eqref{eq:master_equation} a phenomenological term. The latter leaves the electronic part of the equations unaffected, while damping the vibronic excitations towards their thermal equilibrium. It is adapted from Ref.~\cite{Koch2005} to the degenerate AHM. It reads
\begin{equation}
	\LL_\mathrm{rel}\hat{\varrho}^{NmS}
	=
	-\Grel
	\Bigl(\hat{\varrho}^{NmS}-n_B(m\hbar\omega)\sum\limits_{m'}\hat{\varrho}^{Nm' S}\Bigr)
	,
\end{equation}
where $n_B(x)=1/\left(\exp[x/(k_BT)]-1\right)$ is the Bose function and $\Grel$ is a phenomenological relaxation rate.
Assuming that $\Grel \gg \Ga$ is the largest rate in the system, the stationary solution for the GME must be of the form
\begin{equation}
	\hat{\varrho}^{NmS}
	=
	n_B(m\hbar\omega)\sum\limits_{m'}\hat{\varrho}^{Nm' S}
	,
\end{equation}
and we are left with the calculation of the equations of motion for the electronic part of the density matrix $\mat{\varrho}^{NS} = \sum_m\mat{\varrho}^{Nm S}$.
If moreover we concentrate on the regime  $k_B T\ll \hbar\omega$, {favorable} to appreciate the vibrational quantization, only the ground vibronic state is occupied and the equations of motions further simplify to
\begin{align}
	\dot{\varrho}^{00}
	=&
	\sum\limits_{\alpha}\left[
		-4\ga^+ \hat{\varrho}^{00}
		+\ga^- \tr\left\{\Ra\mat{\varrho}^{1\frac{1}{2}}\right\}
	\right]
	,
	\\	\nonumber 
	\dot{\mat{\varrho}}^{1\frac{1}{2}}
	=&
	-\frac{i}{\hbar}
	\left[\HLS^{1\frac{1}{2}},\mat{\varrho}^{1\frac{1}{2}}\right]
	\\	
	&+
	\sum\limits_{\alpha}\left[
		2\ga^+ \Ra \varrho^{00}
		-\frac{1}{2}\ga^- \left\{\Ra,\mat{\varrho}^{1\frac{1}{2}}\right\}
	\right],
\end{align}
with the Lamb shift Hamiltonian	$\mat{H}_{\rm LS}^{1\frac{1}{2}} = \hbar \sum_\alpha \wa \Ra /2$. The 
effective rates and precession frequencies are 
\begin{align}
\label{eq:Rates_T0}
	\ga^\pm
	=&
	\sum\limits_m
	\Ga
	F_{0m}
	f^\pm_\alpha\left(\varepsilon\pm\hbar\omega m\right),
	\\
	\label{eq:Omega_T0}
	\wa
	=&
	\frac{1}{\pi}
	\sum\limits_m
	\Ga
	F_{0m}
	\Bigl[
		p_\alpha\left(\varepsilon-\hbar\omega m\right)
		-
		p_\alpha\left(\varepsilon+U+\hbar\omega m\right)
	\Bigr]
	,
\end{align}
and, as can be derived from Eq.~\eqref{eq:FC_factors}, the Franck-Condon factors reads $F_{0m} = \exp(-g)g^{m}/m!\,$.

\begin{figure}
	\centering
	\includegraphics{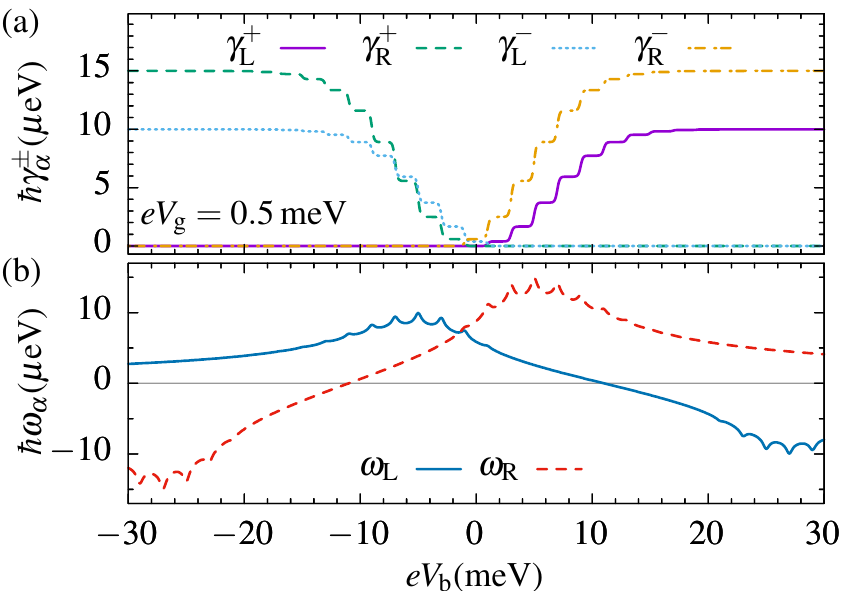}
	\caption{(a) Effective rates and (b) effective precession frequencies as function of the bias voltage.
		The parameters are the same as for Fig.~\ref{fig:stab_diag}, with $eV_{\rm g} = -0.5\,{\rm meV}$}
	\label{fig:effective}
\end{figure}

The effective rates and precession frequencies in Eqs.~\eqref{eq:Rates_T0} and \eqref{eq:Omega_T0} are plotted in Fig.~\ref{fig:effective} as a function of the bias voltage. The step heights in the tunneling rates follow the Poisson distribution $F_{0m}$, as expected from a model with equilibrated vibrons at low temperature. 
The precession frequencies more clearly incorporate the interplay of interference and electron-vibron coupling. By comparison of Fig.~\ref{fig:bias_trace} and Fig.~\ref{fig:effective}, the correspondence between the antiresonance condition of perfect interference blocking for the left to right (right to left) particle current and the vanishing of the precessing frequency $\omega_L = 0$ ($\omega_R = 0$) can be appreciated\cite{Donarini2009}. Moreover, the precession frequencies are locally enhanced every time a new vibronic excitation enters the bias window. In correspondence to such peaks, the current suppression is reduced, since 
the precessing dynamics connects the blocking to the coupled state \cite{Schultz2010,Niklas2017,Donarini2019}. 

In the limit of large bias we can assume $\gL^- = \gR^+ = 0$, and the current can be written as 
\begin{equation}
\label{eq:Current_int}
	I
	\!=\!
	\frac{
		4e\gR \wL^2 \cos(\DP)^2
	}{
		2\gR^2
		+ 2(\wL-\wR)^2
		+ \wL (\wL \gR/\gL + 4\wR\cos(\DP)^2)
	}
	,
\end{equation}
where $\gL^+ = \gL$ and $\gR^- = \gR$, and $\DP = \phi_{\rm R}-\phi_{\rm L}$.
At large negative chemical potential drop, the same expression holds upon exchanging L$\leftrightarrow$R and an overall minus sign.

\begin{figure}
	\centering
	\includegraphics{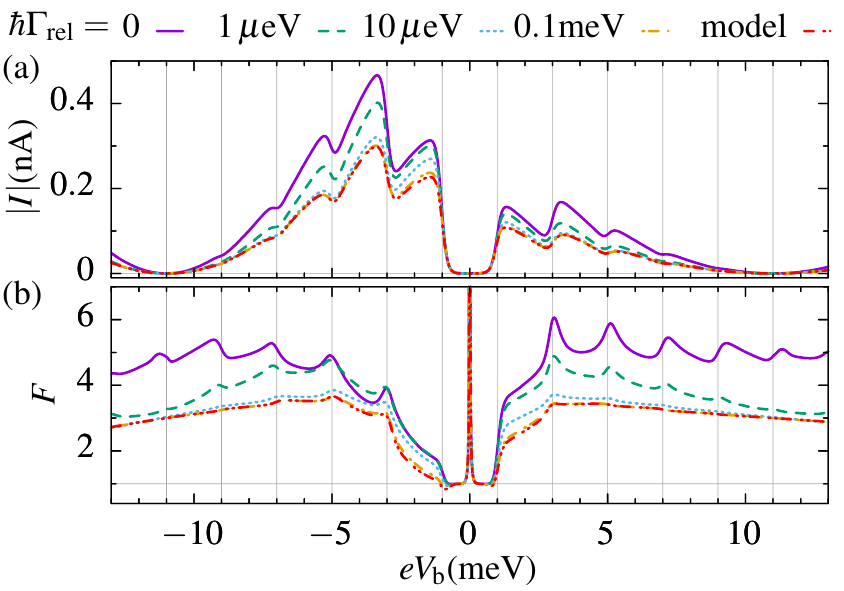}
	\caption{Bias traces of (a) current and (b) Fano factor for different values of the vibronic relaxation rate compared to the analytical high relaxation approximation {(HRA)} {change label accordingly from model to HRA}. The remaining parameters are the same as for Fig.~\ref{fig:stab_diag}.}
	\label{fig:grel}
\end{figure}

It is clear from Eq.~\eqref{eq:Current_int} that the current vanishes as soon as $\wL = 0$ or $\DP = \pi/2$. While the first condition corresponds to a perfect interference blocking\cite{Donarini2010}, the second one has a suggestive interpretation in terms of pseudospin accumulation\cite{Schultz2010} and is completely analogous to the spin-valve set up with anti-parallel polarized leads\cite{Braun2004}. In both cases, though, the pseudospin on the system ends up aligned antiparallel to the drain polarization and the corresponding state is a probability sink.  In Fig.~\ref{fig:grel} we compare the analytical result to numerical calculations for different 
vibronic relaxation rates.
The strength of the relaxation rate should be compared to the renormalized electronic tunneling rates 
$\tilde{\Gamma}_\alpha=\Ga e^{-g}$, giving a ratio $\Gamma_{\rm rel}/\tilde{\Gamma}_\alpha \approx 300$ for the strongest vibronic relaxation and the stronger rate. The current and noise corresponding to the infinite relaxation limit, depicted in red, are asymptotically obtained for every bias condition. Despite the dramatic effects imposed by the vibronic relaxation on the system dynamics, the current remains qualitatively unchanged and suggests to interpret the results in terms of an effective temperature. The vibronic mode would thermalize, even in the absence of a direct dissipative channel, at an effective temperature set by the bias and the strength of the electron-vibron coupling. The Fano factor, though, is more sensitive to the actual dynamics of the system. Its shape changes qualitatively in the high relaxation limit, by rapidly getting flatter and lower, thus revealing the non trivial vibronic contribution to the transport even in this regime of moderate electron-vibron coupling.

\section{Robustness of interference effects}

The robustness of interference effects under vibron induced decoherence has been the focus of theoretical \cite{Ueda2010,Markussen2014,Zhang2015,Penazzi2016} as well as experimental investigation \citep{Ballmann2012,Bessis2016}, although mostly for nanojunction in strong coupling to the electrodes.  
The interference effects in the degenerate Anderson-Holstein model presented so far rely on two necessary conditions: the quasidegeneracy of the orbital states and the absence of independent transport channels connecting both leads, i.e. no basis change can diagonalized simultaneously both tunneling matrices (see Eq.~\eqref{eq:Gmat}). Deviations from these requirements is detrimental for the occurrence of interference. We test here perturbations of different kinds, further proving the robustness of the phenomena described in the previous section. The lifting of the degeneracy by energies comparable or larger {than} the tunneling rates hinders the destructive interference. Several examples of such behavior have been already presented in the literature: the sequential tunneling limit of the conductance canyon\cite{Karlstroem2011} of semiconducting wires can be understood in these terms, as well the quenching of interference in molecules\citep{Darau2009,Schultz2010}, triple quantum dots \citep{Niklas2017} or the absence of two-electron dark states due to exchange interaction in the carbon nanotubes\cite{Donarini2019}. Different electron vibron couplings of the degenerate orbital levels are also destroying the interference blocking \cite{Schultz2010}.

\subsection{Diagonality of tunneling rate matrices}

The most general form of the tunneling rate matrix for a twofold degenerate level reads: 
\begin{equation}
	\boldsymbol{\Gamma}_\alpha = \Gamma_\alpha^0 (\mathbb{1}_2 + {\bf P}_\alpha\cdot \boldsymbol{\sigma})
\end{equation}
where $\boldsymbol{\sigma}$ is the vector of the Pauli matrices, thus identifying the orbital degree of freedom as a pseudospin.
The positivity of the rate matrices is ensured by requiring for the pseudospin polarization vector $|{\bf P}_\alpha| \leq 1$.
By choosing the quantization axis for such a pseudospin perpendicular to the plane spanned by the vectors ${\bf P}_L$ and ${\bf P}_R$ 
one obtains tunneling matrices in the form:
\begin{equation}
	\boldsymbol{\Gamma}_\alpha = \Gamma_\alpha^0 
	\left(
	\begin{array}{cc}
	1 & h_\alpha e^{2i\phi_\alpha}\\
	h_\alpha e^{2i\phi_\alpha} & 1
	\end{array}
	\right) ,
\end{equation}
to be compared with Eq.~\eqref{eq:Gmat} and \eqref{eq:Rmat}. For simplicity, we will assume in our analysis the strength of the pseudospin polarization $h_\alpha$ to be the same in the source and drain lead. 
The factor $0 \leq h \leq 1$ allows one to change between the fully interfering situation ($h=1$) and a diagonal rate matrix ($h = 0$) in which the junction has the same tunneling channels to both leads, and interference is quenched. 
\begin{figure}
	\centering
	\includegraphics{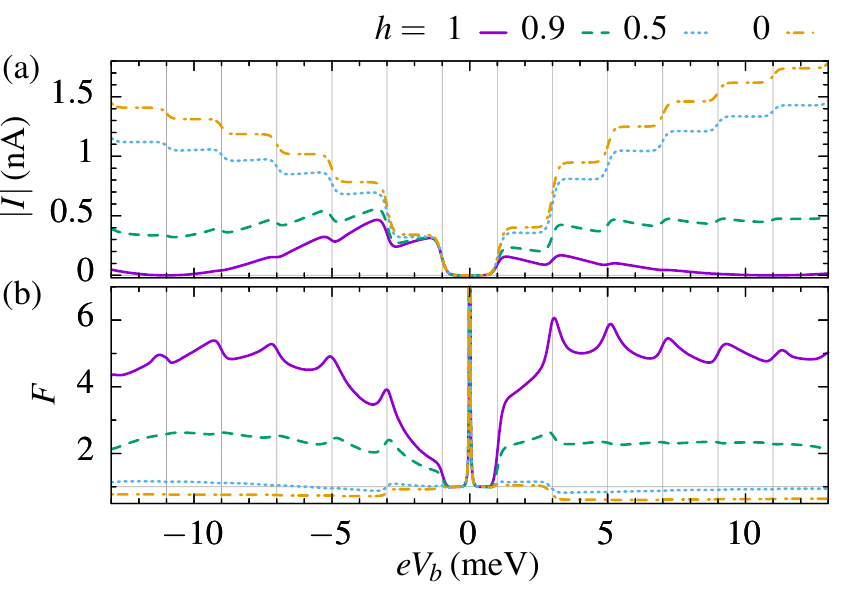}
	\caption{Bias traces of (a) current and (b) Fano factor for different values of the off-diagonal elements of the rate matrix.
		The parameters are the same as for Fig.~\ref{fig:stab_diag}.}
	\label{fig:off-diagonal}
\end{figure}
As one can see at the example of bias traces in Fig.~\ref{fig:off-diagonal}, for fully diagonal rate matrices ($h=0$) the interference vanishes completely and the current consists of regular steps with Poissonian distributed heights. The Fano factor also shows a strong influence on $h$ as it changes from a strongly super-Poissonian behavior to sub-Poissonian behavior, with $F \approx 0.5$, even at large bias voltages.
For any $h>0$ the complete blockade at resonance is disappearing and already at $h \approx 0.5$ NDC is not anymore observable.

\subsection{Full relaxation}

\begin{figure}
	\centering
	\includegraphics{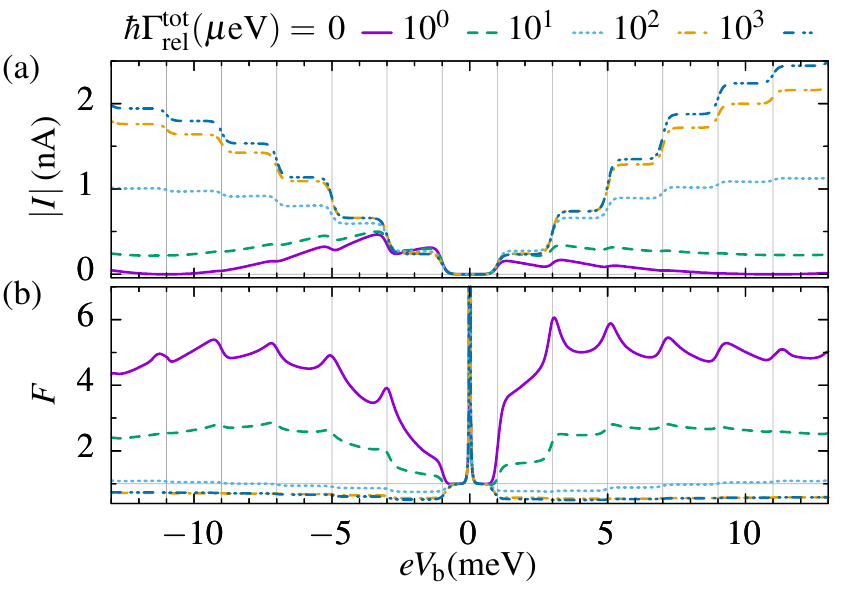}
	\caption{Bias traces of (a) current and (b) Fano factor for different values of the full relaxation rate.
		The parameters are the same as for Fig.~\ref{fig:stab_diag}.}
	\label{fig:grel_full}
\end{figure}

We also test a full relaxation including vibronic as well as electronic degrees of freedom via the term
\begin{equation}
\LL_\mathrm{rel}^\mathrm{tot}
\hat{\varrho}^{NmS}
=
-\Grel^\mathrm{tot}\left(
\hat{\varrho}^{NmS}
-
\hat{\varrho}^{NmS}_\mathrm{th}
\sum\limits_{m'\ell}\hat{\varrho}^{Nm' S}_{\ell\ell}
\right)
,
\end{equation}
where in the thermal density matrix $\varrho^{0mS}_{\rm th} = n_B(m\hbar\omega)$ and $\mat{\varrho}^{1mS}_{\rm th} = n_B(m\hbar\omega)\mat{1}_2/2$ is diagonal in the orbital degree of freedom.
Figure ~\ref{fig:grel_full} displays the current and Fano factor for different values of the total relaxation rate.
This type of relaxation destroys the interference as soon as the total relaxation rate becomes of the order of the tunneling rates. For larger total relaxation rates the step like current behavior of vibronic systems is recovered together with sub-Poissonian statistics, as previously reported in the literature \citep{Haupt2006}.

Destructive interference occurs, in principle, for all phase differences $\DP \neq 0$ between the left and the right tunneling processes. However, the smaller $\DP$ the more sensitive is the system to independent processes of relaxation towards the thermal solution which bring the current to its incoherent limit. This phenomenon, already shown in absence of vibrational degrees of freedom (see the supplementary information of Ref.~ \onlinecite{Donarini2019}) is not qualitatively modified in the degenerate Anderson-Holstein model by the presence of the mechanical degrees of freedom.

\section{Conclusions}

Motivated by experiments on suspended {CNT quantum dots}  \citep{Sapmaz2006,Leturcq2009,Huettel2009,Stiller2018} and by the recent observation, again on CNTs, of coherent population trapping, we have studied the interplay of quantum interference and vibronic degrees of freedom in the transport through an interacting  nanojunction. 
The model of choice is the degenerate Anderson-Holstein model \citep{Schultz2010}, in which the single interacting level is replaced by a pair of orbitally degenerate ones. The different tunneling phases of these levels towards the two leads support interference phenomena and thus coherent population trapping. We calculate for this system both the current and the current noise,  expressed using the Fano factor as indicator of the transport dynamics. 

The current voltage characteristics appear as a superposition of vibron assisted tunneling and quantum interference. The first is  responsible for  the step-like increase of the current whenever a new vibronic excitation enters the transport window. Interference, instead, modulates the current on a larger energy scale: it induces the negative differential conductance following each current step and, ultimately, the complete current suppression at the antiresonance conditions $e\Vg + \eta e\Vb = -U/2$ and $e\Vg + (\eta-1) e\Vb = -U/2$.

The precession dynamics between the coupled and the decoupled states plays a central role in the understanding of the degenerate Anderson-Holstein model. The analytic expression for the current derived in the strong relaxation limit clearly emphasizes this aspect through a pronounced dependence on the precession frequency. The latter, in turn, depends on the bias and the gate voltage, showing how all-electrical control of the coherent dynamics of an interacting electromechanical system can be obtained.  

Finally, the robustness of the effects presented so far is analyzed.  Beyond the lifting of the electronic degeneracy, already extensively considered in the literature \cite{Darau2009,Schultz2010,Karlstroem2011,Niklas2017,Donarini2019} particular emphasis has been given to the degree of pseudospin polarization and the strength of electronic as well as vibronic relaxation.

The degenerate Anderson-Holstein model, with its interplay of interference and vibron assisted tunneling, represents an interesting and {rich} playground for the understanding of the transport characteristics of interacting  electromechanical systems with a degenerate electronic spectrum. In particular, the combined analysis of its current and current noise characteristics suggests an alternative interpretation of controversial measurements on suspended CNT quantum dots, which emphasises the role of Lamb shift corrections and pseudospin precession for these nanojunctions.

\section{Acknowledgement}

The financial support of the Deutsche Forschungsgemeinschaft under the research program CRC 1277 project B02 is {acknowledged}.

\appendix

\section{Liouville space}

This effective model can be written in Liouville space where the equation takes the form $\dot{\hat{\varrho}}=\mathcal{L}\hat{\varrho}$ and where $\hat{\varrho} = ( \hat{\varrho}^{00}, \hat{\varrho}^{1\frac{1}{2}}_{-\ell-\ell}, \hat{\varrho}^{1\frac{1}{2}}_{\ell\ell}, \hat{\varrho}^{1\frac{1}{2}}_{-\ell\ell}, \hat{\varrho}^{1\frac{1}{2}}_{\ell-\ell} )^\intercal$. The Liouvillian and current operators read
\begin{widetext}
\begin{align}
	\LL
	&=
	\begin{pmatrix}
		-4 \gamma^+ & 
			\gamma^- &
				\gamma^- &
					\tilde{\gamma}^{-*} &
						\tilde{\gamma}^- \\
		2 \gamma^+ &
			-\gamma^- &
				0 &
					\frac{1}{2}(i\tilde{\omega}^* - \tilde{\gamma}^{-*}) &
						\frac{1}{2}(-i\tilde{\omega} - \tilde{\gamma}^-) \\
		2 \gamma^+ &
			0 &
				-\gamma^- &
					\frac{1}{2}(-i\tilde{\omega}^* - \tilde{\gamma}^{-*}) &
						\frac{1}{2}(i\tilde{\omega} - \tilde{\gamma}^-) \\
		2 \tilde{\gamma}^+ &
			\frac{1}{2}(i\tilde{\omega} - \tilde{\gamma}^-) &
				\frac{1}{2}(-i\tilde{\omega} - \tilde{\gamma}^-) &
					-\gamma^- &
						0 \\
		2 \tilde{\gamma}^{+*} &
			\frac{1}{2}(-i\tilde{\omega}^* - \tilde{\gamma}^{-*})	&
				\frac{1}{2}(i\tilde{\omega}^* - \tilde{\gamma}^{-*}) &
					0 &
						-\gamma^-
	\end{pmatrix}
	, \\
	\JP
	&=
	\begin{pmatrix}
		0 & \gR^- & \gR^- & e^{i\DP}\gR^- & e^{-i\DP}\gR^- \\
		0 & 0 & 0 & 0 & 0 \\
		0 & 0 & 0 & 0 & 0 \\
		0 & 0 & 0 & 0 & 0 \\
		0 & 0 & 0 & 0 & 0
	\end{pmatrix},
	\quad
	\JM
	=
	\begin{pmatrix}
		0 & 0 & 0 & 0 & 0 \\
		2\gR^+ & 0 & 0 & 0 & 0 \\
		2\gR^+ & 0 & 0 & 0 & 0 \\
		2e^{-i\DP}\gR^+ & 0 & 0 & 0 & 0 \\
		2e^{i\DP}\gR^+ & 0 & 0 & 0 & 0
	\end{pmatrix}
\end{align}
with $\phi_R-\phi_L=\DP$, $\tilde{\gamma}^\pm = e^{i\DP}\gL^\pm + e^{-i\DP}\gR^\pm$, $\tilde{\omega} = e^{i\DP}\wL + e^{-i\DP}\wR$ and $\gamma^\pm = \gL^\pm + \gR^\pm$.
\end{widetext}

\bibliography{bib}

\end{document}